%% file: Main.tex
 \documentclass[manuscript,sigconf]{acmart}
\input{meta/preamble}

\begin{document}

\input{meta/title}

\input{meta/authors}

\input{meta/abstract}

\input{meta/cencepts_keywords}


\maketitle

\section{Introduction}
\input{sections/intro}

\section{Applications and experimental setup}
\label{sect:list_of_applications_and_setup}
\input{sections/setup}

\section{Experimental results} \label{sect:experimental_results}
\input{sections/exp-results}



\begin{acks}
\input{sections/acks}
\end{acks}

 \bibliographystyle{ACM-Reference-Format}
 \input{references.tex}


\end{document}

%% file: meta/preamble.tex
\copyrightyear{2021}
\acmYear{2021}
\setcopyright{acmcopyright}\acmConference[PEARC '21]{Practice and Experience in Advanced Research Computing}{July 18--22, 2021}{Boston, MA, USA}
\acmBooktitle{Practice and Experience in Advanced Research Computing (PEARC '21), July 18--22, 2021, Boston, MA, USA}
\acmPrice{15.00}
\acmDOI{10.1145/3437359.3465592}
\acmISBN{978-1-4503-8292-2/21/07}

\usepackage{booktabs}
\usepackage{xcolor}

%% file: meta/title.tex
\title{Comparing the behavior of OpenMP Implementations with various Applications on two different Fujitsu A64FX platforms}

%% file: meta/authors.tex
\author{Benjamin Michalowicz}
\email{benjamin.michalowicz@stonybrook.edu}
\affiliation{%
  \institution{Stony Brook University}
  \state{New York}
  \country{USA}
}

\author{Eric Raut}
\affiliation{%
  \institution{Stony Brook University}
  \city{New York}
  \country{USA}}
\email{eric.raut@stonybrook.edu}

\author{Yan Kang}
\affiliation{%
  \institution{Stony Brook University}
  \city{New York}
  \country{USA}}
\email{yan.kang@stonybrook.edu}

\author{Tony Curtis}
\affiliation{%
  \institution{Stony Brook University}
  \city{New York}
  \country{USA}}
\email{anthony.curtis@stonybrook.edu}

\author{Barbara Chapman}
\affiliation{%
  \institution{Stony Brook University}
  \city{New York}
  \country{USA}}
\email{barbara.chapman@stonybrook.edu}

\author{Dossay Oryspayev}
\affiliation{%
  \institution{Brookhaven National Laboratory}
  \city{New York}
  \country{USA}}
\email{doryspaye@bnl.gov}

\renewcommand{\shortauthors}{B.T. Michalowicz, E. Raut, Y. Kang et al.}

%% file: meta/abstract.tex
\begin{abstract}
The development of the A64FX processor by Fujitsu has been a massive innovation in vectorized processors and led to Fugaku: the current world's fastest supercomputer.
We use a variety of tools
to analyze the
behavior and performance of several OpenMP applications with different compilers, and
how these applications scale on the different A64FX processors on clusters at Stony Brook University and RIKEN.
\end{abstract}

%% file: meta/cencepts_keywords.tex
\ccsdesc[500]{Computer systems organization~HPC Architecture}
\ccsdesc[300]{High Performance Computing~OpenMP}
\ccsdesc{High Performance Computing~Parallel Programming}
\ccsdesc[100]{Computer systems organization~Compiler Toolchains}

\keywords{OpenMP, High Performance Computing, Ookami, A64FX, Fujitsu, Fugaku}

%% file: sections/intro.tex
The introduction of the A64FX processor by Fujitsu
has sparked an innovation in vectorized processors and the birth of Fugaku: the current world's-fastest supercomputer~\footnote{November, 2020, list of {\color{blue} \url{https://top500.org}}}. The A64FX chip also brings an unprecedented co-design approach, impressive performance, and energy-awareness that puts it at the top position on all $5$ major HPC benchmarks. In this (short) paper, we analyze the OpenMP~\cite{OpenMP} shared-memory/parallel programming model, from how it scales on the  A64FX -- and its variants on Ookami and Fugaku, see Section \ref{sect:list_of_applications_and_setup} -- to performance across different compiler toolchains.\footnote{This work is in progress and here we present a partial list of results we so far have due to page limitations.}

The A64FX processor~\cite{RyohiOkazaki2020,Sato2020} is the processor
used in the Fugaku supercomputer,
enabled by
the Japanese FLAGSHIP 2020 project as a co-design between RIKEN and Fujitsu. Currently, Fugaku is ranked number $1$ on both the Top500 and HPCG lists. A64FX is a general-purpose processor based on the Armv8.2-A specification~\cite{Sato2020} and
has $48$ compute cores, divided into four core memory groups (CMGs)
with $12$ cores, and $2-4$ cores dedicated to OS communications. 

OpenMP is a directive-based standard for parallel programming on shared memory systems. Its ease of use makes OpenMP very attractive for obtaining efficient parallel versions of serial programs. The programmer can use compiler directives, library routines, and environment variables to write parallel programs for shared memory systems in Fortran and C/C++.

%% file: sections/setup.tex
\subsection{List of Applications}

\begin{itemize}
    \item Minimod~\cite{Meng2020} - a seismic modeling mini-application that solves the acoustic wave equation.
    Minimod is
    used to study the performance of emerging compilers and runtimes for HPC. An OpenMP task-based
    version ~\cite{Raut2020} is used in this paper.
    \item PENNANT~\cite{LANLPennant} - a mesh physics mini-application for advanced architecture research, with the mesh size determined by input sets. PENNANT is dominated by pointer chasing.

    It can be run solely with MPI or in a hybrid MPI+OpenMP setup. It uses OpenMP's static loop-scheduling, and makes use of gather/scatter to send data to and from the root to other ranks, and reductions to consolidate partial results.
    \item SWIM~\footnote{{\color{blue} \url{https://www.spec.org/cpu2000/CFP2000/171.swim/docs/171.swim.html}}} - a Fortran OpenMP weather forecasting program designed for testing current performance of supercomputers. Like PENNANT, SWIM also uses static scheduling in OpenMP loops.
\end{itemize}

\subsection{Systems and Compilers}\label{SystemCompiler}

\subsubsection*{Fugaku}

The world's fastest supercomputer, located at The RIKEN Center for Computational Science in Japan~\cite{R-CCS}.
Its processor is the Fujitsu A64FX, and has a proprietary interconnect called Tofu, configured as a $6$D torus. The cluster became operational in $2020$ and enters production usage in $2021$. Fugaku provides Fujitsu's
native and cross-compilers alongside various versions of GNU compilers. Its underlying Tofu-D interconnect is not used in these experiments -- only intranode performance is measured in our experiments. In addition, its stripped-down, lightweight, customized Linux kernel allows users to further enhance their applications' performance through the use of various environment variables not defined in standard Linux distributions.

\subsubsection*{Ookami}

A new cluster installed at Stony Brook University (SBU) in late summer
2020. It has $174$ compute nodes, with another $2$ for debugging/experimentation. Ookami was funded through an NSF grant~\cite{NSF-1927880} as the first A64FX cluster outside of Japan. Ookami uses a non-blocking HDR $200$ switching fabric via $9$ $40$-port Mellanox Infiniband switches in a $2$-level tree. Each node, similar to Fugaku's compute nodes, currently has $32$GB of high-bandwidth memory. Our experiments do not use these switches in its intranode experiments. In addition, it runs a standard CentOS Linux distribution and its A64FX processors do not contain the extra cores that come in the "full" FX1000 variant of the chip. Table~\ref{tab:table1} shows the compilers used.

\begin{table}[h!]
  \begin{center}
    \begin{tabular}{|c|c|c|} \hline
      \multicolumn{1}{|c}{} & \multicolumn{2}{|c|}{\textbf{Versions}}\\
      \hline
      \textbf{Compiler Family} & \textbf{Fugaku} & \textbf{Ookami} \\
      \hline
      ARM & - & $20.3$\\
      Cray & - & $10.0.1$\\ 
      Fujitsu & $4.3.0\text{a}$ & -\\
      GCC & $8.3.1$, $10.2.1$ & $8.3.1$, $10.2.1$, $11.0.0$\\
      LLVM & $11.0.0$ & $11.0.0$, $12.0.0$\\
      \hline
    \end{tabular}
    \caption{Compilers of Fugaku and Ookami.}
    \label{tab:table1}
  \end{center}
\end{table}

\vspace{-0.1in}

\subsection{Runtime Environment}
Each benchmark was run on one compute node with one MPI rank to avoid shared memory operations that occur with two or more processes, and over-subscription of threads to cores, which might result in degraded performance. Threads are bound to cores using the \texttt{{OMP\_PLACES}} environment variable with its semantics \texttt{{start\_core:num\_cores}}. This allows threads to be assigned to specific cores (e.g., Thread $0$ is assigned to Core $0$) as well as
splitting threads among specific CMGs. 

\subsection{Compiler options}\label{Optimizations}
For each compiler mentioned in
Section \ref{SystemCompiler},
we enabled specific flags, 
maximizing thread optimization and SVE instructions, and minimizing execution-time while maintaining correctness. We also enabled fine-tuning for the A64FX processor, where possible. The flags are listed for each compiler/group\footnote{For any GNU and LLVM compiler: If compiling directly on an A64FX node, use \texttt{{-mcpu=native}} instead.} as shown in Table~\ref{tab:table2}.

\begin{table}[h!]
\begin{center}
    \begin{tabular}{ |c|c|l| }
        \hline
        \textbf{Compiler} & \textbf{Flags} \\
        \hline
        Cray & \parbox[t]{4.3cm}{\texttt{-homp -hvector3 -hthread3}} \\
        \hline
        GCC & \parbox[t]{4.3cm}{\texttt{-mcpu=a64fx \\-Ofast -fopenmp}}\\ 
        \hline
        LLVM & \parbox[t]{4.3cm}{\texttt{-mcpu=a64fx \\-Ofast -fopenmp}} \\ 
        \hline 
        Fujitsu-Traditional & \parbox[t]{4.3cm}{\texttt{-Nnoclang -Nlibomp -O3 -KSVE,fast,openmp,ARMV8\_2\_A}} \\ 
        \hline
        Fujitsu-LLVM & \parbox[t]{4.3cm}{\texttt{-Nclang -Nlibomp -Ofast \\-Kfast,openmp -mcpu=a64fx+sve}}\\
        \hline
    \end{tabular}

    \caption{Flags used for each compiler.}
    \label{tab:table2}
\end{center}
\end{table}

%% file: sections/exp-results.tex
We analyzed
runtime and relative speed-up using OpenMP with different compiler classes -- Cray, ARM, GNU, Fujitsu, and LLVM; and different versions among each class where applicable. Each set of results is drawn from running our programs on specified inputs five times per specified OpenMP thread count (one from 1, 2, 4, 8, 12, 16, 24, 32, 36, and 48) and obtaining the arithmetic mean for each set. Note that for each graph, the x-axis refers to the number of OpenMP threads, ranging from 1 to 48.

\subsection{Ookami}{\label{OokamiRes}}

For each application, we made graphs of the three compilers we deemed "best in class": of Ookami's compilers across the GCC, ARM, and Cray collections, binaries compiled from these gave the best runtime performance and speedup: GCC 10.2.0, ARM 20.1.3, and Cray 10.0.1,
with the results for the other compilers explained in the following subsections.

\subsubsection{PENNANT}

For this application we focus on the \textbf{LeblancBig} input, which
fits in the $32$GB of on-chip memory while also being a non-trivial input size.

\begin{figure}[ht]
    \centering
	\includegraphics[width=.8\linewidth]{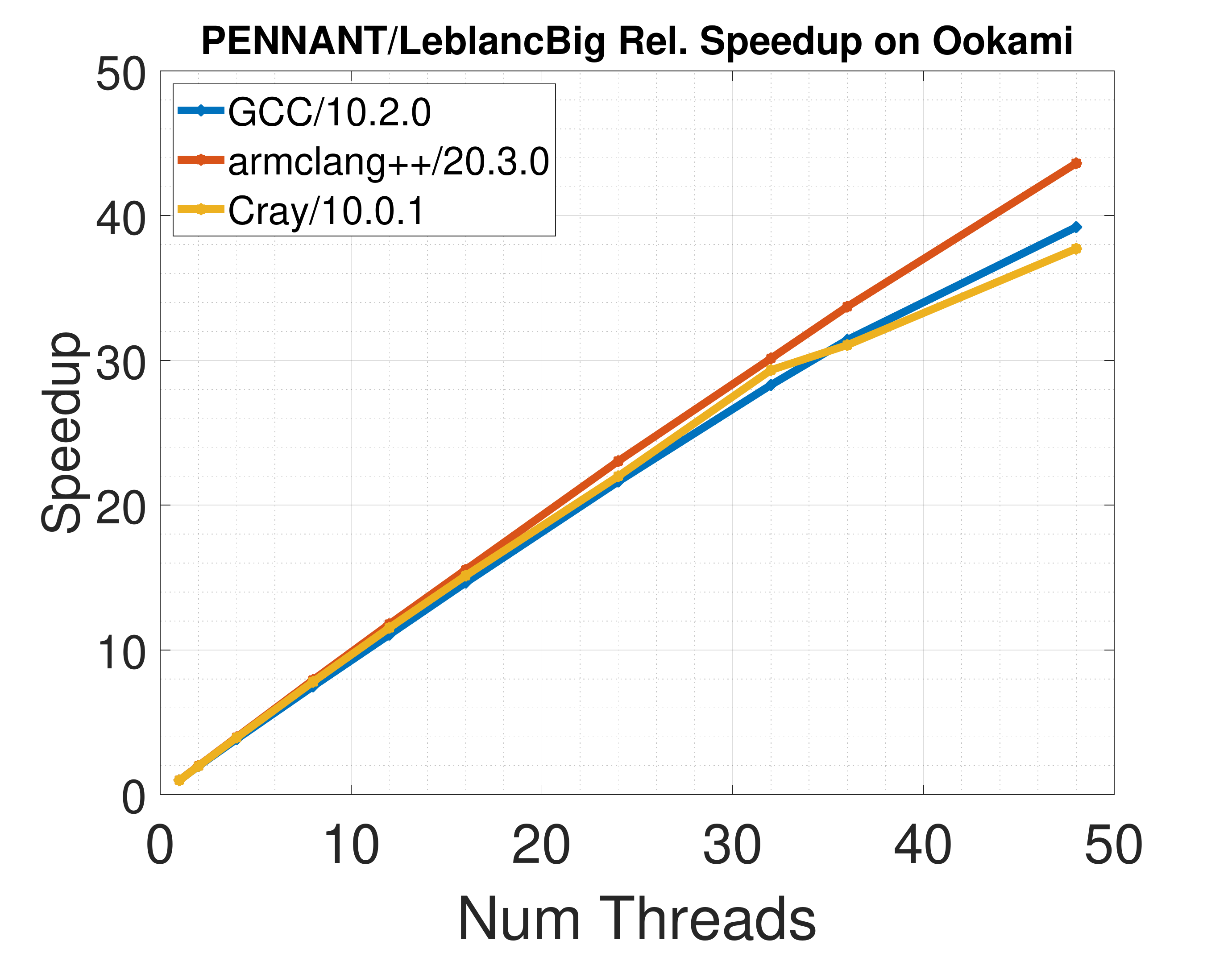}
	\caption{LeblancBig Speedup Comparison}
	\label{fig:leblancbig-speedup}
\end{figure}

In Figure \ref{fig:leblancbig-speedup}, we show the relative speedup observed between each of the three compilers, measured by the runtime $T_{thread\_value}$ compared to $T_{1\_OpenMP\_Thread}$. While the \texttt{armclang}-compiled code returns the slowest overall runtimes, it gives the most linear relative speed-up, with the Cray compilers having the smallest relative speed-up from being able to quickly saturate the on-chip memory bandwidth at the start of execution.

Profiling the \textbf{LeblancBig} input with CrayPat~\cite{CrayPat} and ARM Forge~\cite{ArmForge} on Ookami, we noticed that different values for \texttt{{OMP\_WAIT\_POLICY}} (active or passive) resulted in substantially different behaviors. An active wait policy resulted in PENNANT spending $66.3$\% of its runtime in OpenMP regions. Conversely, the passive wait policy results in only $17.8$\% of \textbf{LeblancBig}'s runtime inside OpenMP regions. The difference in time spent between computation and synchronization of threads is proportional to the requested thread count, with very little time (under $30$\%)  used for thread synchronization.

\subsubsection{SWIM}

\begin{figure}[ht]
    \centering
	\includegraphics[width=0.7\linewidth]{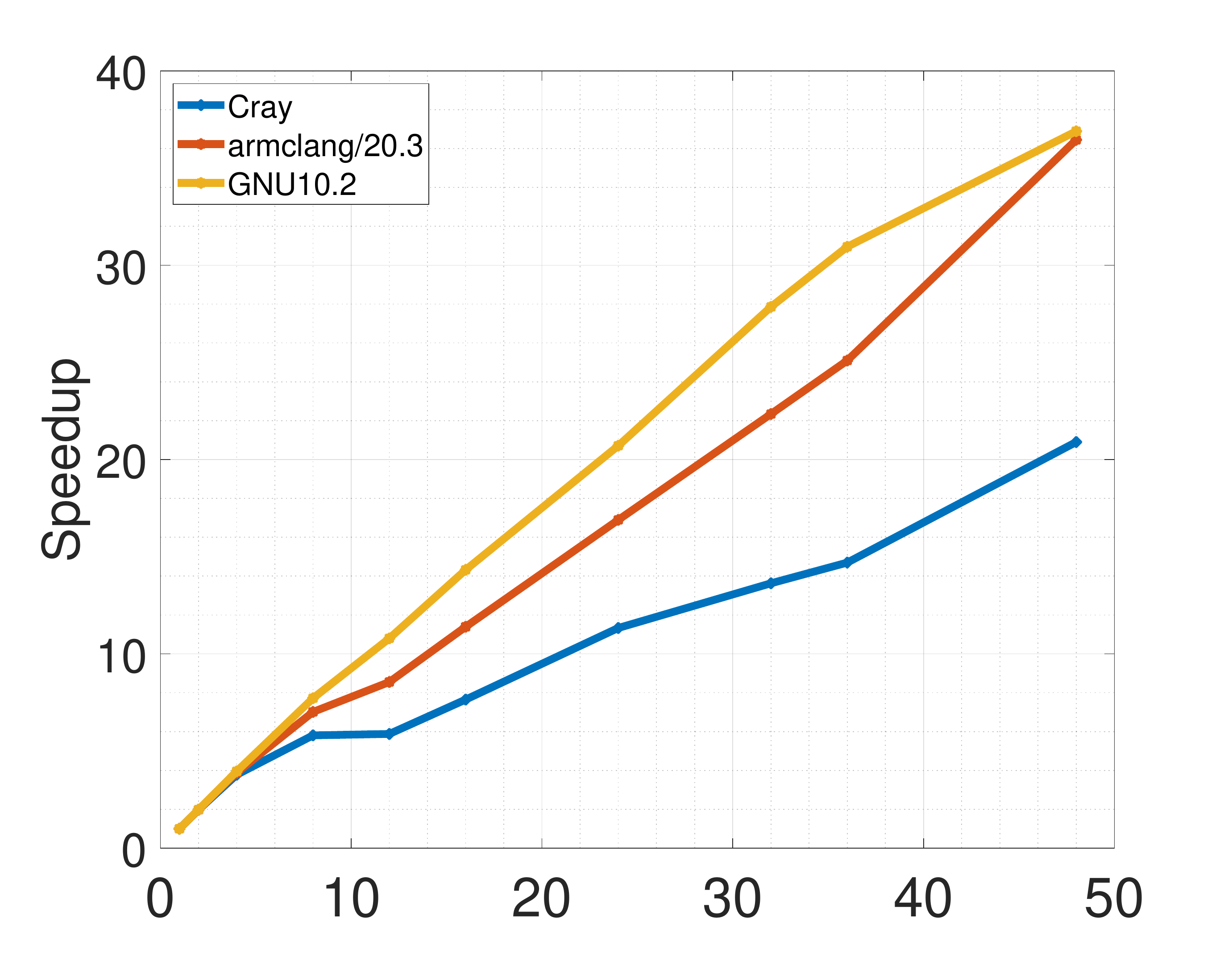}
	\caption{SWIM: speed-up results}
	\label{fig:swim-speedup}
\end{figure}

The default input for SWIM is \texttt{{swim.ref.in}}, which sets up a $7701x7701$ matrix running for $3000$ iterations. In our experiments, we tested $7$ different compiler versions, but to avoid clutter and data overlap, we have chosen $3$ representatives from the various compiler families: GNU, ARM's LLVM-based compiler, and Cray. We present speed-up results as shown in Figure~\ref{fig:swim-speedup}.

As shown in Figure~\ref{fig:swim-speedup}, we can see that among all of the compilers, the GNU compiler seems to have the greatest speed-up. $48$ threads achieve a $37$x speed-up over $1$ thread.
On the other hand, the Cray compiler has a much more moderate speed-up rate. The greatest speed-up is about $16$ times between $1$ and $48$ OpenMP threads. 

With profiling tools ARM MAP and CrayPat~\cite{CrayPat} on Ookami, SWIM spends $70.2$\% of its runtime on OpenMP regions, which is understandable since it is a purely OpenMP benchmark. OpenMP generates a small amount of overhead: $28.5$\% was seen with this particular run.

\subsubsection{Minimod}

Two different OpenMP configurations of Minimod (see~\cite{Raut2020} for details) were evaluated:
\begin{itemize}

    \item Loop xy: Grid is blocked in $x$ (largest-stride) and $y$ dimensions. An OpenMP parallel for loop is applied to the $2$-D loop nest over x-y blocks. (A \texttt{{collapse(2)}} is used to combine the two loops).
    \item Tasks xy: Grid is blocked in $x$ and $y$ dimensions. Each x-y block is a task using OpenMP's tasking directive. OpenMP's \texttt{{depend}} clause is used to manage dependencies between time steps.
\end{itemize}

\begin{figure}[ht]
    \centering
	\includegraphics[width=0.7\linewidth]{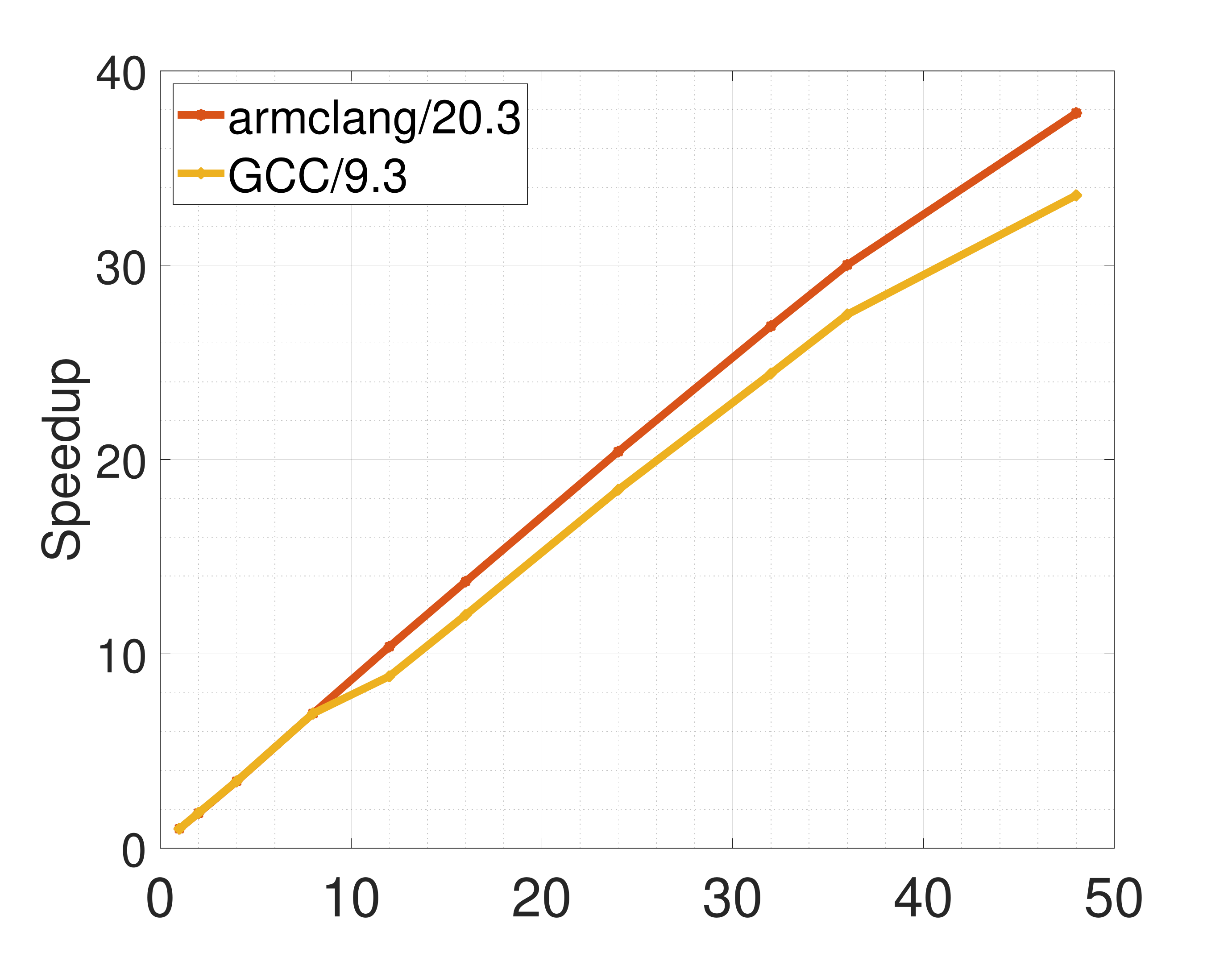}
	\caption{Minimod: speedup plot (tasks xy) for each compiler}
	\label{fig:minimod-speedup-tasksxy}
\end{figure}

In both cases a grid size of $512^3$ was used. Minimod speedups are shown for the tasks-xy configuration in Figure~\ref{fig:minimod-speedup-tasksxy}.
In general, for the task-based configuration, LLVM tends to outperform GCC (particularly at higher thread counts), although the total runtimes for this grid size are quite similar. In the loop-xy configuration, GCC performs slightly better than LLVM in terms of total runtime, and the speedups are similar between the two compilers.

Profiling using the Arm Forge Performance Report tool, we find that both configurations are entirely compute-bound, and both have a high number of stalled cycles ($76.5$\% and $80.7$\% of cycles for loop-xy and tasks-xy configurations, respectively), indicating that the application is memory-bound. This makes the HBM2 memory of the A64FX processor potentially advantageous for
these applications.


\subsection{Fugaku}\label{FugakuRes}
 The Fujitsu compiler has two backends (traditional; LLVM), so we can compare performance and thread-scaling between them.
 In this section, we will break down and explain our results on Fugaku comparing results between the GNU and Fujitsu compilers. Per the experiments in Section~\ref{sect:list_of_applications_and_setup}, we ran each application $5$ times and took the average of the runtimes. 

\subsubsection{PENNANT}\label{PennantFugaku}
The Fujitsu compiler gave the longest recorded runtimes for the \textbf{LeblancBig} input.
In particular, the single-threaded runtimes for both inputs had surprisingly large standard deviations ($107$ seconds as opposed to a fraction of $1$ second). Both versions of the GNU compilers on Fugaku were still competitive with Ookami. We noticed that the traditional back-end options for the Fujitsu compiler, compared to the LLVM-backend (see section~\ref{Optimizations}), took substantially longer in smaller-threaded runs. Profiling \textbf{LeblancBig} shows that the traditional Fujitsu back-end generally results in a less efficient execution compared to the LLVM back-end. In particular, both back-ends are faster at $24$ OpenMP threads ($181$ seconds with LLVM, $186$ with traditional) than at $48$ threads ($233$ and $236$ seconds, respectively).

\begin{figure}[ht]
    \centering
	\includegraphics[width=0.7\linewidth]{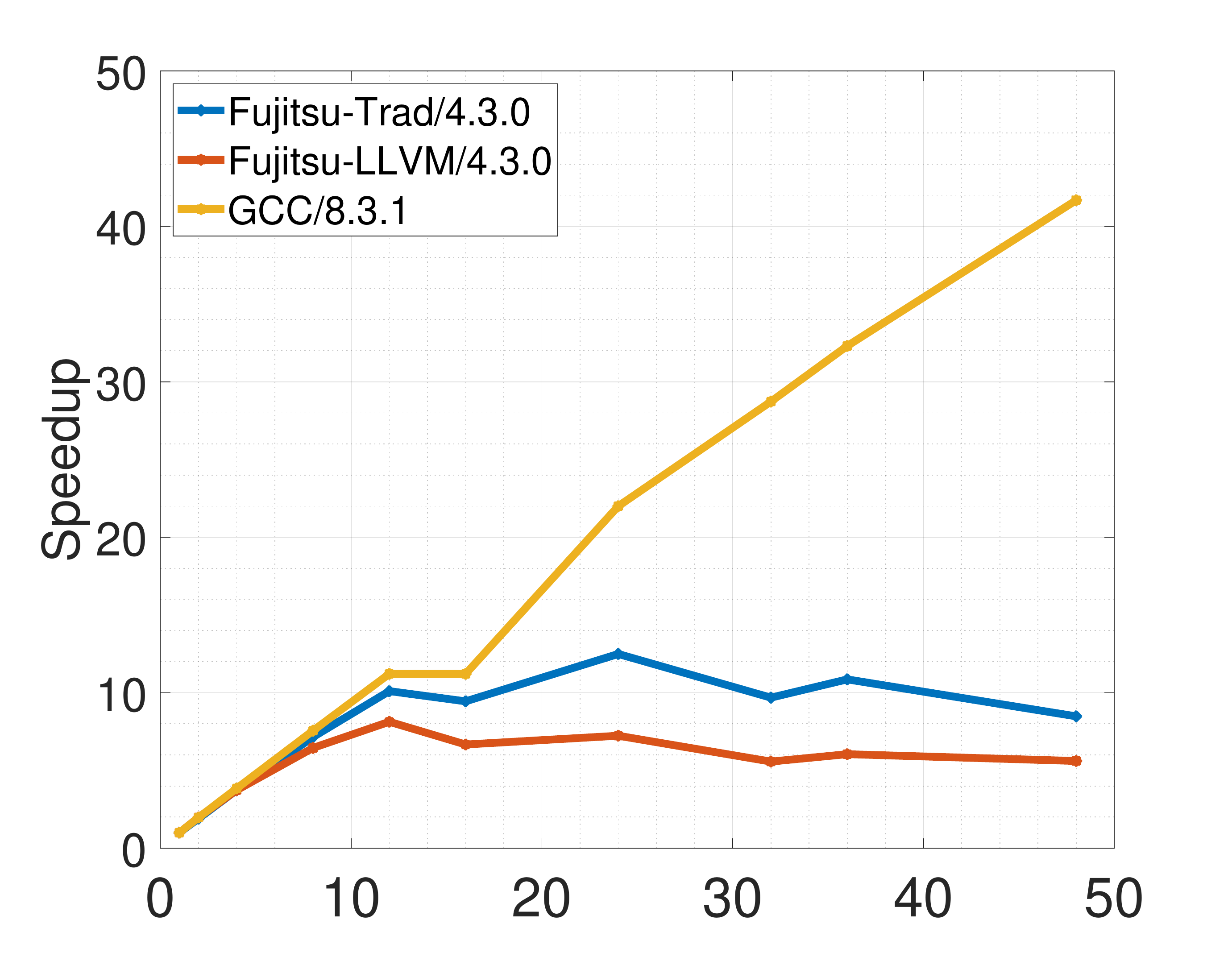}
	\caption{LeblancBig/Fugaku Speedup}
	\label{LeblancBig-Fugaku-Speed}
\end{figure}

Similarly, this results in observed reduced speedup for the Fujitsu compiler, especially after reaching $12$ threads placed in $1$ CMG (see Figure~\ref{LeblancBig-Fugaku-Speed})\footnote{While not the focus of our study, this observation would be resolved by utilizing PENNANT's MPI features in addition to OpenMP.}.

\subsubsection{SWIM}
\begin{figure}[ht]
    \centering
	\includegraphics[width=0.7\linewidth]{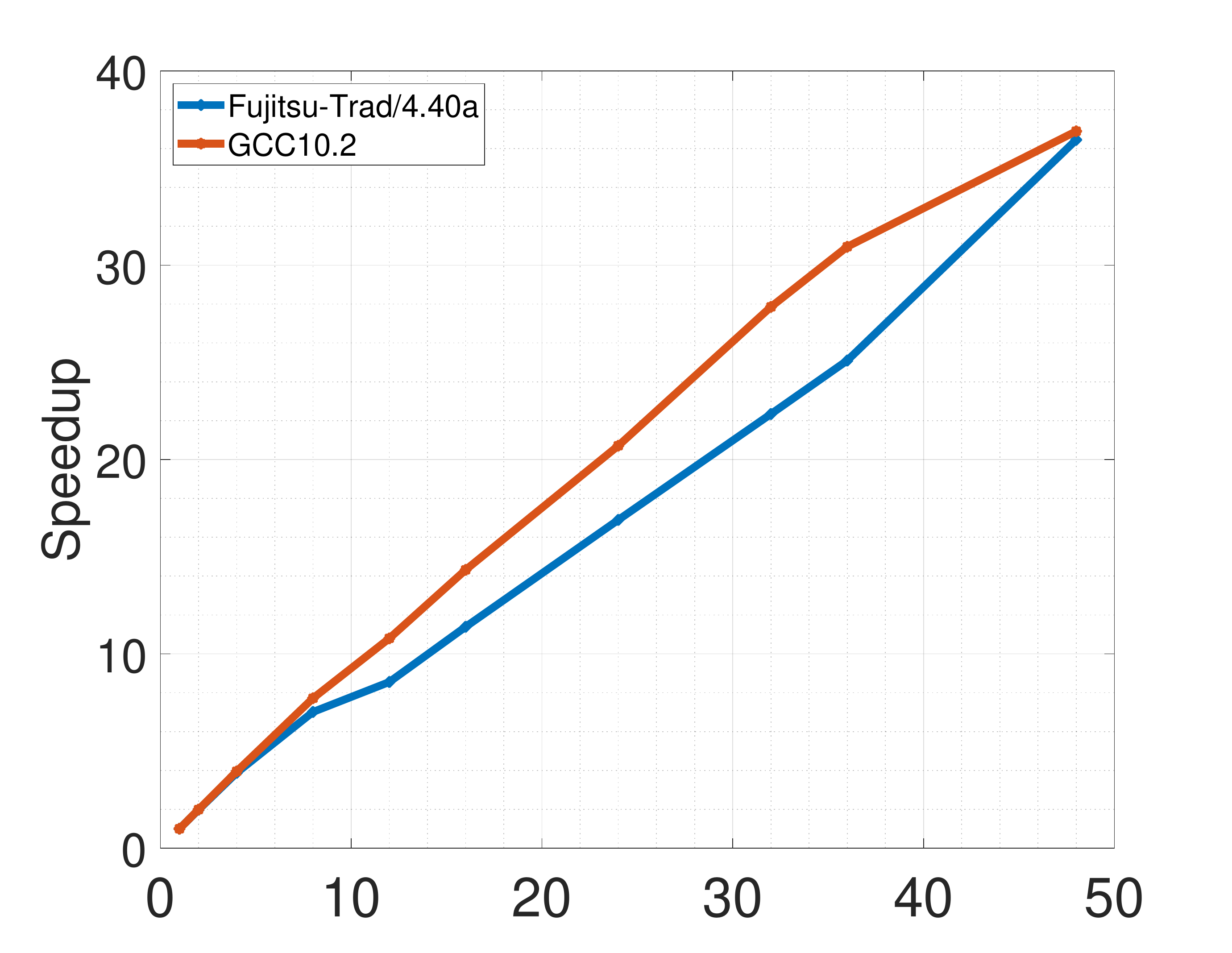}
	\caption{SWIM: Speedup Plot}
	\label{fig:swim-fugaku-Speedup}
\end{figure}

On Fugaku, the Fujitsu and GCC ($10.2.1$) compilers were used to test SWIM's capabilities. Compared to results reported in section~\ref{PennantFugaku}, SWIM ran significantly faster when compiled with Fujitsu than with Cray on Ookami.
As shown in Figure~\ref{fig:swim-fugaku-Speedup}, the GNU compiler has a greater speed-up than Fujitsu. $36$ threads achieve a $32$x speed-up over $1$ thread. The Fujitsu compiler only obtained a $25$x speed-up with $36$ threads over $1$ thread. The Fugaku-based runs show a drop in relative speed-up starting at $8$ OpenMP threads before leveling out at $12$ threads, which could be explained by insufficient OpenMP optimization of the Fujitsu compiler.

On Fugaku, SWIM has shown a much better performance than all other applications tested previously on Ookami. It has achieved a performance of over $31$ GFLOPS, which correlates well with the high SVE operation rate ($99.9983$\%).

\subsubsection{Minimod}

\begin{figure}[ht]
    \centering
	\includegraphics[width=0.7\linewidth]{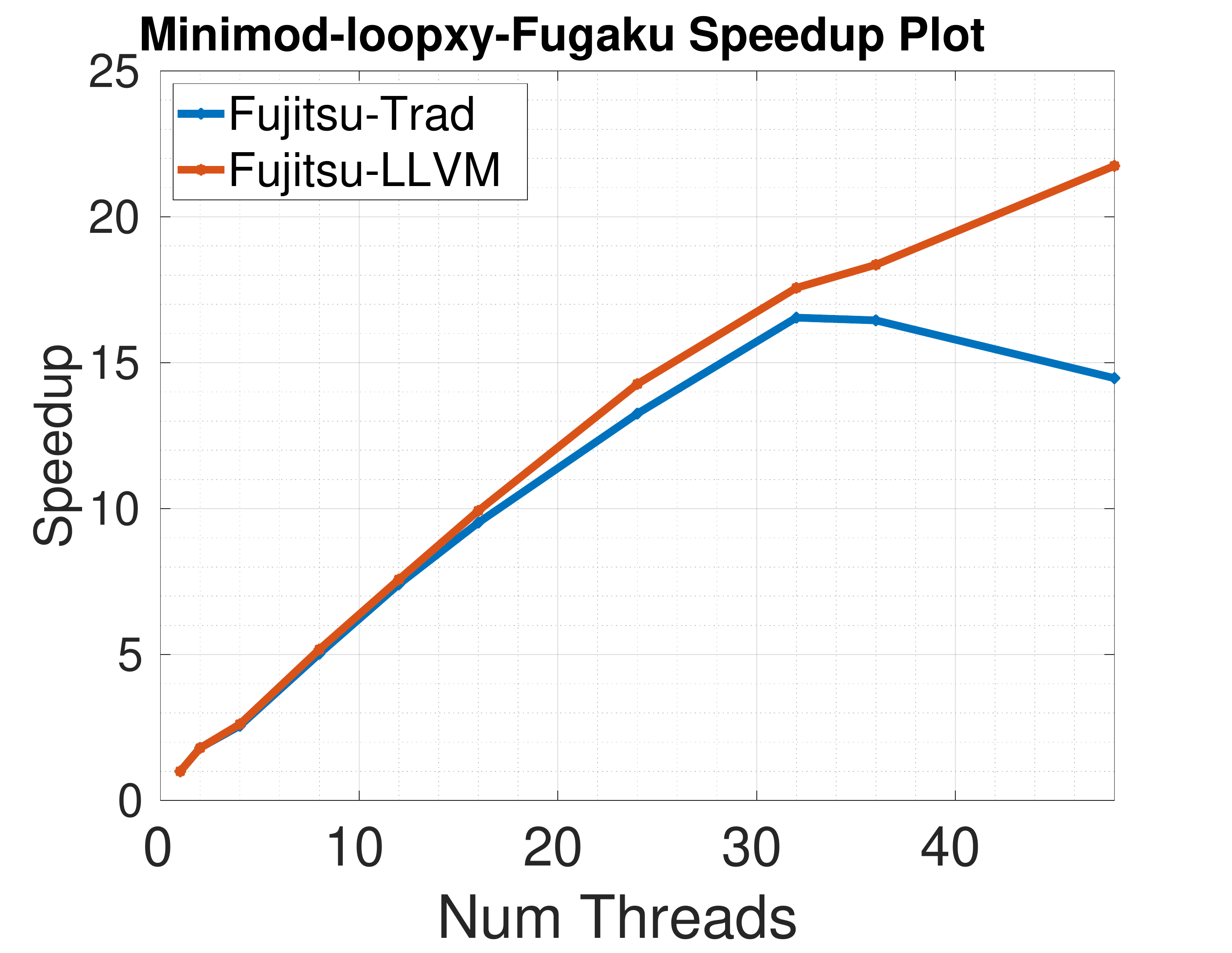}
	\caption{Minimod: speedup plot (loop xy) for each compiler on Fugaku}
	\label{fig:minimod-speedup-loopxy-fugaku}
\end{figure}

Figure~\ref{fig:minimod-speedup-loopxy-fugaku} shows the speedup result for the ``loopxy'' configuration of Minimod. The Fujitsu-Traditional compiler suffers a significant slowdown at higher thread counts, although the speedups for both compilers are significantly worse than the compilers evaluated on Ookami.

The Fujitsu-Trad compiler failed to run the tasks-xy configuration, presumably because of a lack of support for the OpenMP \texttt{depend} clause of the task directive. We are still working on evaluating with other compilers on Fugaku.

%% file: sections/acks.tex
We would like to thank the NSF for supporting the Ookami cluster, and the ability to research the Fujitsu A64FX processor, through grant OAC 1927880. We would like to thank the Riken Center for Computational Science for providing us with access to Fugaku and to conduct research on it. We would also like to thank Stony Brook University and the Institute for Advanced Computational Science for providing the resources to allow us to conduct our studies on Ookami.

%% file: references.tex

%% file: Main.bbl
\begin{thebibliography}{10}


\ifx \showCODEN    \undefined \def \showCODEN     #1{\unskip}     \fi
\ifx \showDOI      \undefined \def \showDOI       #1{#1}\fi
\ifx \showISBNx    \undefined \def \showISBNx     #1{\unskip}     \fi
\ifx \showISBNxiii \undefined \def \showISBNxiii  #1{\unskip}     \fi
\ifx \showISSN     \undefined \def \showISSN      #1{\unskip}     \fi
\ifx \showLCCN     \undefined \def \showLCCN      #1{\unskip}     \fi
\ifx \shownote     \undefined \def \shownote      #1{#1}          \fi
\ifx \showarticletitle \undefined \def \showarticletitle #1{#1}   \fi
\ifx \showURL      \undefined \def \showURL       {\relax}        \fi
\providecommand\bibfield[2]{#2}
\providecommand\bibinfo[2]{#2}
\providecommand\natexlab[1]{#1}
\providecommand\showeprint[2][]{arXiv:#2}

\bibitem[\protect\citeauthoryear{ARM}{ARM}{[n.d.]}]%
        {ArmForge}
\bibfield{author}{\bibinfo{person}{ARM}.} \bibinfo{year}{[n.d.]}\natexlab{}.
\newblock \bibinfo{booktitle}{\emph{ARM Forge Documentation}}.
\newblock
\urldef\tempurl%
\url{https://developer.arm.com/documentation/101136/2021/Performance-Reports}
\showURL{%
\tempurl}


\bibitem[\protect\citeauthoryear{Enterprise}{Enterprise}{[n.d.]}]%
        {CrayPat}
\bibfield{author}{\bibinfo{person}{Hewlett~Packard Enterprise}.}
  \bibinfo{year}{[n.d.]}\natexlab{}.
\newblock \bibinfo{booktitle}{\emph{CrayPat Documentation}}.
\newblock
\urldef\tempurl%
\url{https://pubs.cray.com/bundle/HPE_Performance_Analysis_Tools_User_Guide_S-8014_2012/page/CrayPat_Runtime_Environment.html}
\showURL{%
\tempurl}


\bibitem[\protect\citeauthoryear{Ferenbaugh}{Ferenbaugh}{[n.d.]}]%
        {LANLPennant}
\bibfield{author}{\bibinfo{person}{Charles~R. Ferenbaugh}.}
  \bibinfo{year}{[n.d.]}\natexlab{}.
\newblock \bibinfo{booktitle}{\emph{PENNANT: An Unstructured Mesh Mini-App for
  Advanced Architecture Research}}.
\newblock
\urldef\tempurl%
\url{https://www.osti.gov/biblio/1079561-pennant-unstructured-mesh-mini-app-advanced-architecture-research}
\showURL{%
\tempurl}


\bibitem[\protect\citeauthoryear{Meng, Atle, Calandra, and Araya-Polo}{Meng
  et~al\mbox{.}}{2020}]%
        {Meng2020}
\bibfield{author}{\bibinfo{person}{Jie Meng}, \bibinfo{person}{Andreas Atle},
  \bibinfo{person}{Henri Calandra}, {and} \bibinfo{person}{Mauricio
  Araya-Polo}.} \bibinfo{year}{2020}\natexlab{}.
\newblock \bibinfo{title}{Minimod: A Finite Difference solver for Seismic
  Modeling}.
\newblock
\newblock
\showeprint[arxiv]{2007.06048}~[cs.DC]


\bibitem[\protect\citeauthoryear{NSF}{NSF}{[n.d.]}]%
        {NSF-1927880}
\bibfield{author}{\bibinfo{person}{NSF}.} \bibinfo{year}{[n.d.]}\natexlab{}.
\newblock \bibinfo{title}{Ookami: A high-productivity path to frontiers of
  scientific discovery enabled by exascale system technologies}.
\newblock
  \bibinfo{howpublished}{\url{{https://www.nsf.gov/awardsearch/showAward?AWD_ID=1927880}}}.
\newblock


\bibitem[\protect\citeauthoryear{Okazaki, Tabata, Sakashita, Kitamura, Takagi,
  Sakata, Ishibashi, Nakamura, and Ajima}{Okazaki et~al\mbox{.}}{2020}]%
        {RyohiOkazaki2020}
\bibfield{author}{\bibinfo{person}{Ryohi Okazaki}, \bibinfo{person}{Takekazu
  Tabata}, \bibinfo{person}{Sota Sakashita}, \bibinfo{person}{Kenichi
  Kitamura}, \bibinfo{person}{Noriko Takagi}, \bibinfo{person}{Hideki Sakata},
  \bibinfo{person}{Takeshi Ishibashi}, \bibinfo{person}{Takeo Nakamura}, {and}
  \bibinfo{person}{Yuichiro Ajima}.} \bibinfo{year}{2020}\natexlab{}.
\newblock \showarticletitle{Supercomputer Fugaku CPU A64FX Realizing High
  Performance, High-Density Packaging, and Low Power Consumption}.
\newblock
  \bibinfo{howpublished}{\url{https://www.fujitsu.com/global/about/resources/publications/technicalreview/2020-03/article03.html}}.
\newblock \bibinfo{journal}{\emph{Fujitsu Rechnical Review}}
  (\bibinfo{date}{Nov.} \bibinfo{year}{2020}).
\newblock
\urldef\tempurl%
\url{https://www.fujitsu.com/global/about/resources/publications/technicalreview/2020-03/article03.html}
\showURL{%
\tempurl}


\bibitem[\protect\citeauthoryear{OpenMP-ARB}{OpenMP-ARB}{[n.d.]}]%
        {OpenMP}
\bibfield{author}{\bibinfo{person}{OpenMP-ARB}.}
  \bibinfo{year}{[n.d.]}\natexlab{}.
\newblock \bibinfo{booktitle}{\emph{OpenMP Website}}.
\newblock
\urldef\tempurl%
\url{https://www.openmp.org}
\showURL{%
\tempurl}


\bibitem[\protect\citeauthoryear{Raut, Meng, Araya-Polo, and Chapman}{Raut
  et~al\mbox{.}}{2020}]%
        {Raut2020}
\bibfield{author}{\bibinfo{person}{Eric Raut}, \bibinfo{person}{Jie Meng},
  \bibinfo{person}{Mauricio Araya-Polo}, {and} \bibinfo{person}{Barbara
  Chapman}.} \bibinfo{year}{2020}\natexlab{}.
\newblock \showarticletitle{Evaluating Performance of OpenMP Tasks in a Seismic
  Stencil Application}. In \bibinfo{booktitle}{\emph{OpenMP: Portable
  Multi-Level Parallelism on Modern Systems}},
  \bibfield{editor}{\bibinfo{person}{Kent Milfeld}, \bibinfo{person}{Bronis~R.
  de~Supinski}, \bibinfo{person}{Lars Koesterke}, {and} \bibinfo{person}{Jannis
  Klinkenberg}} (Eds.). \bibinfo{publisher}{Springer International Publishing},
  \bibinfo{address}{Cham}, \bibinfo{pages}{67--81}.
\newblock
\showISBNx{978-3-030-58144-2}
\urldef\tempurl%
\url{https://doi.org/10.1007/978-3-030-58144-2_5}
\showDOI{\tempurl}


\bibitem[\protect\citeauthoryear{RIKEN}{RIKEN}{[n.d.]}]%
        {R-CCS}
\bibfield{author}{\bibinfo{person}{RIKEN}.} \bibinfo{year}{[n.d.]}\natexlab{}.
\newblock \bibinfo{booktitle}{\emph{Fugaku Project}}.
\newblock
\urldef\tempurl%
\url{\\https://www.r-ccs.riken.jp/en/fugaku/project}
\showURL{%
\tempurl}


\bibitem[\protect\citeauthoryear{Sato, Ishikawa, Tomita, Kodama, Odajima,
  Tsuji, Yashiro, Aoki, Shida, Miyoshi, Hirai, Furuya, Asato, Morita, and
  Shimizu}{Sato et~al\mbox{.}}{2020}]%
        {Sato2020}
\bibfield{author}{\bibinfo{person}{Mitsuhisa Sato}, \bibinfo{person}{Yutaka
  Ishikawa}, \bibinfo{person}{Hirofumi Tomita}, \bibinfo{person}{Yuetsu
  Kodama}, \bibinfo{person}{Tetsuya Odajima}, \bibinfo{person}{Miwako Tsuji},
  \bibinfo{person}{Hisashi Yashiro}, \bibinfo{person}{Masaki Aoki},
  \bibinfo{person}{Naoyuki Shida}, \bibinfo{person}{Ikuo Miyoshi},
  \bibinfo{person}{Kouichi Hirai}, \bibinfo{person}{Atsushi Furuya},
  \bibinfo{person}{Akira Asato}, \bibinfo{person}{Kuniki Morita}, {and}
  \bibinfo{person}{Toshiyuki Shimizu}.} \bibinfo{year}{2020}\natexlab{}.
\newblock \showarticletitle{Co-Design for A64FX Manycore Processor and
  "Fugaku"}. In \bibinfo{booktitle}{\emph{Proceedings of the International
  Conference for High Performance Computing, Networking, Storage and Analysis}}
  (Atlanta, Georgia) \emph{(\bibinfo{series}{SC '20})}.
  \bibinfo{publisher}{IEEE Press}, Article \bibinfo{articleno}{47},
  \bibinfo{numpages}{15}~pages.
\newblock
\showISBNx{9781728199986}


\end{thebibliography}
